\documentclass[12pt]{article}
\usepackage{amsfonts}
\usepackage{amsmath,amssymb}
\usepackage{fancybox}
\usepackage[dvips]{color}
\usepackage[dvips]{graphicx}

\numberwithin{equation}{section}

\begin{document}  
\titlepage
\setcounter{page}{0}
\null \hfill Preprint TU-898  \\[3em]

\begin{center}
{\large\bf Boundary state analysis on 
the equivalence of T-duality and Nahm transformation 
in superstring theory }\\[2em] 
\end{center}

\begin{center}
{T. Asakawa\footnote{
e-mail: asakawa@tuhep.phys.tohoku.ac.jp}, U. Carow-Watamura\footnote{
e-mail: ursula@tuhep.phys.tohoku.ac.jp}, Y. Teshima\footnote{
e-mail: teshima@tuhep.phys.tohoku.ac.jp}, and S. Watamura\footnote{
e-mail: watamura@tuhep.phys.tohoku.ac.jp}}\\[3em] 

Department of Physics \\
Graduate School of Science \\
Tohoku University \\
Aoba-ku, Sendai 980-8578, Japan \\ [2ex]

\end{center}

~~~~


\def\rc#1{{\color{red}#1}}
\def\be{\begin{equation}}
\def\ee{\end{equation}}
\def\bea{\begin{eqnarray}}
\def\eea{\end{eqnarray}}
\def\CA{{\cal A}}\def\CB{{\cal B}}\def\CD{{\cal D}} 
\def\CE{{\cal E}}\def\CN{{\cal N}}\def\CO{{\cal O}}
\def\CH{{\cal H}}
\def\CP{{\cal P}}
\def\complex{\mathbb{C}}
\def\real{\mathbb{R}}
\def\matrix#1#2{
\begin{array}{#1}#2\end{array}}
\def\Tr#1{{\rm T}\!{\rm r}\{#1\}}

\def\integer{\mathbb{Z}}
\def\id{{\bf1}}
\def\half{{1\over2}}
\def\norm#1{|\!|#1 |\!|}
\def\Bra#1{\langle #1|}
\def\Ket#1{|#1 \rangle}
\def\braket#1{\langle #1 \rangle}

\def\sla#1{\setbox0=\hbox{$#1$} 
\dimen0=\wd0 
\setbox1=\hbox{/} \dimen1=\wd1 
\ifdim\dimen0>\dimen1 
\rlap{\hbox to \dimen0{\hfil/\hfil}} 
#1 
\else 
\rlap{\hbox to \dimen1{\hfil$#1$\hfil}} 
/ 
\fi} %

\begin{abstract}

We investigated the equivalence of the T-duality for a bound state of $D2$ and 
$D0$-branes 
with the Nahm transformation of the corresponding gauge theory 
on a $2$-dimensional torus, 
using the boundary state analysis in superstring theory. 
In contrast to the case of a $4$-dimensional torus, it changes a sign 
in a topological charge, 
which seems puzzling when regarded as a $D$-brane charge.
Nevertheless, it is shown that it agrees with the T-duality of the boundary state,
including a minus sign.
We reformulated boundary states in the RR-sector using a new 
representation of zeromodes,
and show that the RR-coupling is invariant under the T-duality.
Finally, the T-duality invariance at the level of the Chern-Simon coupling 
is shown by deriving the Buscher rule for the RR potentials, known as the 'Hori formula', 
including the correct sign.

\end{abstract}

\eject

\section{Introduction}

T-duality in string theory is a genuine stringy symmetry, which does not exist in a particle theory.
In the compactified space a string does not distinguish whether it propagates on a spacetime with a circular dimension of radius $R$ or ${\alpha'\over R}$, 
a property discussed already more than 20 years ago \cite{Kikka, Sakai}. 
The investigations during the last decade show that string theory should possess 
an extremely rich symmetry of this type, including the relations 
among various dimensional D-branes.

At present, there are many approaches throughout in the literature, discussing T-duality in various contexts and developed rather independently: An early formulation of the T-duality was given in \cite{Busch}. Then, discussions from the world sheet point of view of string theory and the investigation of string states, including the boundary states, followed 
\cite{Giveon,T-dual,open T-dual,Kugo,Dai}. 
The T-duality was also formulated at the level of the low energy effective theory. 
In the closed string sector the transformation rules in the supergravity, known as the Buscher rules, were formulated in \cite{RR Buscher} and were further investigated in 
\cite{Zumino,Fukuma,Hassan,Hori}.
More recently, the T-duality with nontrivial $H$-flux was investigated 
by taking the global structure of spacetime into account, see e.g. \cite{Bouwknegt2010}. 
To include the open string sector, the T-duality has to be extended to a formulation including D-branes. 
Under T-duality, the effective theory of the D-branes relates the theories in 
various dimensions, and 
the effect of the T-duality on the Chern-Simons term \cite{Zumino,Green,Minasian,Sundell}, and on matrix models \cite{Taylor} were discussed.
It was also recognized \cite{Hori} that a certain gauge theory duality, i.e. a
Nahm transformation \cite{Nahm, ADHM}, relates to 
the T-duality of the bound states of the branes,
with subsequent arguments \cite{Kaji,Terashima}.
On the other hand, in order to make contact with phenomenology, 
compactification scenarios for the
extra dimensions into rather complicated backgrounds such as Calabi-Yau spaces \cite{Candel} or certain generalized spaces are considered. Generalizations of the Riemann geometry are proposed and a fiberwise T-duality of bundles or sheaves are discussed. In these pictures, the D-branes are described by coherent sheaves, and the T-duality transformations are considered to be Fourier-Mukai transforms along certain fibers (see e.g. \cite{Andre}). 

Our aim here is to analyze the compatibility among
the results on T-duality found in the various approaches, 
i.e. worldsheet, supergravity and gauge theory, discussed above.
The boundary state description of $D$-branes is 
an appropriate framework for such a purpose, since a boundary state 
can be defined for $D$-branes with a non-trivial gauge bundle on it, and its coupling to 
various closed string states is also easily estimated. 
This became possible due to the 
recent progress achieved in constructing the boundary state with nontrivial gauge bundle 
on a compact worldvolume (\cite{Duo,DiVecc} and references therein). 
We treat here the $2$-dimensional torus using the boundary state formulation of a $D2/D0$-brane system and examine its compatibility with the corresponding Nahm-Fourier-Mukai duality.

The Nahm transformation was originally proposed for the case of a $4$-dimensional gauge theory with the aim to construct the monopole solution \cite{Nahm} and instanton solution \cite{ADHM}. The case of the instanton solution on the $4$-dimensional torus is well understood: The Nahm transformation gives a map from the $k$ instanton solution of a $SU(N)$ gauge theory on a torus $T^4$ to the $N$ instanton solution of the $SU(k)$ gauge theory on its dual 4-torus $\tilde T^4$ \cite{BB,Schenk}. 

The role of the Nahm transformation in the string theory becomes transparent with the 
observation, along the line of \cite{Douglas,Diaco}, 
that at low energies the theory of $N$ $D4$-branes wrapping a $4$-dimensional torus $T^4$ corresponds to a $U(N)$ gauge theory on a bunch of $N$ $D4$-branes including $k$ $D0$ branes and corresponds to a $k$-instanton solution in this gauge theory. 
The T-dual map $T^4\rightarrow \tilde T^4$ transforms this $ND4/kD0$ brane system into a system of $kD4/ND0$ branes. 
The compatibility of T-duality and the Nahm transformation in this case was observed 
in \cite{Hori} by using the probe analysis.

Since the case of dimension $4$ is rather special, 
we want to check this situation for tori of dimensions other than $4k$. 
The Nahm transformation of $T^2$ case has been previously considered in \cite{Kaji},
however, important sign factors needed for consistency were missing there. 
As a result the compatibility with T-duality could not be justified.
In this paper, we first describe the $2$-dimensional Nahm transformation in detail, 
by following \cite{Kaji} mainly, but correcting some formulae. 
Unlike to the $T^4$ case where an anti-selfdual (ASD) gauge field is transformed into an ASD gauge field on the dual torus, in the 2-dimensional case the Nahm transformation maps a uniform magnetic flux $F_{\mu\nu}$ of $T^2$ into that of the dual 2-torus $\tilde{F}_{\mu\nu}$. 
These states correspond to the BPS bound states of $D2$-branes and $D0$-branes, where $D0$-branes are dissolved inside $D2$-branes as a uniform flux.

Since the gauge field configuration is rather simple, we can explicitly construct the corresponding boundary states and compare the T-duality and the Nahm transformation directly on the string level. 
For the bosonic string sector, 
we will use the formulation of boundary states with a constant magnetic flux 
constructed in \cite{Duo, DiVecc}.
As shown in \cite{Duo}, even this simple case, T-duality transformation is accompanied 
by an appropriate cocycle factor.
Including such a cocycle factor, we will establish the precise agreement of the T-duality transformation of the boundary state with the Nahm transformation in the bosonic string theory.

This result at first sight seems odd because the minus sign in the 
dual magnetic flux suggests
an anti-$D0$-brane rather than a $D0$-brane.
In order to estimate the correct RR-charges carried by the dual boundary state, 
we have to extend the boundary states to the fermionic part.
Therefore, we construct the boundary state in the superstring theory. For this end, we will give a new representation of the RR-zeromodes.
It gives the same amplitude as the standard representation after the GSO projection,
but as we will see, it is suitable when considering the T-duality transformation.
In this setup, it is easy to show the invariance of the RR-coupling of the boundary state,
namely, the Nahm transformation gives the desired RR-charges in the dual theory. 
We will also derive the T-duality rule for RR-potentials, known as Buscher rule or Hori formula,
and show the invariance of the Chern-Simons term in terms of low energy fields.
We emphasize that the overall signs appearing in these formulae are automatically 
implied by the string theory T-duality and are important to maintain the invariance.

The plan of this paper is as follows. In section 2, we will 
describe the Nahm transformation in detail.
In section 3, a brief review of the construction of boundary states on the torus 
following \cite{Duo,DiVecc} is given and the explicit form for our gauge 
configuration is derived. 
Then we perform the T-duality transformation of this boundary state.
In section 4, we construct the fermionic part of the boundary state 
using a representation which simplifies the pairing with the RR-state,  
and compare the T-duality and Nahm transformation. 
Then the coupling to RR-potentials are investigated.
Section 5 gives conclusions and discussion.


\section{Nahm transformation on the 2-dimensional torus}
\label{sec:Nahm}

We consider $U(N)$ bundle $E$ over the 2-dimensional torus $T^2$ with bundle connection  denoted by $A_{\mu}$. 
To define the Nahm transformation which maps the $U(N)$ bundle $E$ over a torus $T^2$ to its dual one $\tilde{E}$ over $\tilde{T}^2$, we introduce the Poincar$\rm \acute{e}$ bundle ${\cal P} \rightarrow T^2\times \tilde{T}^2$, which is a complex line bundle with 
curvature ${1\over2\pi}dx^\mu \wedge d\tilde{x}_\mu$, where 
$x^\mu$ and $\tilde{x}_\mu$ are coordinates on $T^2,\tilde{T}^2$ respectively. 
We also define the projection $\pi: T^2\times \tilde{T}^2 \rightarrow T^2$, (and correspondingly for the dual torus). 

Roughly speaking, the Nahm transform of $E$ is defined by the following process: 
take the pull-back to $T^2\times \tilde{T}^2$, tensor with the Poincar\'e bundle ${\cal P}$, and then project to define a bundle $\tilde{E}$ on $\tilde{T}^2$ as the index bundle of the Dirac operator, as we will now describe.
Considering the bundle ${\cal E}=\pi^*E\otimes {\cal P}$ over $T^2\times \tilde T^2$, 
we can 
obtain a family of 
covariant derivatives $D_{\tilde{x}\,\mu}(x)=\partial_\mu - i A_\mu(x) -i{\tilde{x}_\mu\over 2\pi}$ by restricting the bundle to 
${\cal E}_{\tilde x}=\pi^*E\otimes {\cal P}|_{T^2\times\{ \tilde{x}\}}$.
The covariant derivative is then parametrized by $\tilde{x}$. 
Given a Dirac operator, which maps a section of 
the spin bundle $S^{\pm}$ coupled to $E$ on $T^2$ to the opposite chirality:
\be
\CD^{\pm}: \Gamma(T^2, S^{\mp}\otimes E) \rightarrow \Gamma(T^2, S^{\pm}\otimes E)\ ,
\ee 
the corresponding Dirac operator coupled with ${\cal E}$ is decomposed into ${\cal D}^+_{\tilde{x}},{\cal D}^-_{\tilde{x}}$
as
\begin{eqnarray}
	\sla D_{\tilde{x}} & = &
    \gamma^\mu D_{\tilde{x}\,\mu}\nonumber\\
    & = &
      \left(
      \begin{array}{cc}
        0 & {\cal{D}}^+_{\tilde{x}}\\
        {\cal{D}}^-_{\tilde{x}} & 0
      \end{array}
      \right)~,
\end{eqnarray}
where $\gamma_\mu=-i\sigma_\mu$ with the Pauli 
matrices $\sigma_\mu,~\mu=1,2$, and 
\begin{eqnarray}
&&	{\cal D}^+_{\tilde{x}} 
    = -i(\partial_1 -i A_1 -i{\tilde{x}_1\over 2\pi})
      - (\partial_2 -i A_2 -i{\tilde{x}_2\over2\pi})~,\nonumber\\
&&	{\cal D}^-_{\tilde{x}}
    = -i(\partial_1 -i A_1 -i{\tilde{x}_1\over2\pi}) 
      + (\partial_2 -i A_2 -i{\tilde{x}_2\over2\pi})~.
\end{eqnarray}

From now on we consider a bundle $E$ with a positive first Chern number, $C_1(E)=k>0$
and we look for the Dirac zero modes.
In this case, the equation 
${\cal D}^-_{\tilde{x}}\psi = 0$ has no normalized solutions for the left-handed component $\psi \in \Gamma(T^2, \pi^* S^{+}\otimes {\cal E}_{\tilde x})$, it follows from the index 
theorem that for the right-handed spinor $\psi \in \Gamma(T^2, \pi^* S^{-}\otimes {\cal E}_{\tilde x})$, 
\begin{eqnarray}
  {\cal D}^+_{\tilde{x}}\psi = 0			\label{Dirac}
\end{eqnarray}
has $k$ normalized solutions $\psi^p(p=1,\cdots,k)$, that depend on the parameter $\tilde{x}$. They span the vector space $H_{\tilde x}\simeq \complex^k$ and can be collected into a 
$N\times k$ matrix of zero-modes $\Psi$. 

The Hilbert space $\CH=L^2 (T^2, \pi^* S^{-}\otimes {\cal E}_{\tilde x})$ of the fermions $\psi$ will be decomposed into space of the zeromodes $H_{\tilde x}$ 
 and its complement. 
Thus,
with appropriate conditions for completeness and finite norm $\norm{\psi}^2<\infty$, 
we can define a projection $P_{\tilde x}$ to the zeromode subspace 
\be
P_{\tilde x}=\sum_p \Ket{\psi^p}\Bra{\psi^p}=\Ket{\Psi}\Bra{\Psi} ,
\ee
which acts on the vector $\chi\in \CH$ as
\be
P_{\tilde x}\chi\equiv \sum_p \psi^p\braket{\psi^p|\chi}=\sum_p\psi^p\int_{T^2}d^2x{{\psi^p}^{\dagger}\chi}.
\ee

The Nahm transform ${\cal N}$ of $(E,A)$ over a torus $T^2$ is a pair $(\tilde{E},\tilde{A})$ of a vector bundle and a connection, where $\tilde{E}$ is a smooth vector bundle over the dual torus $\tilde{T}^2$, 
 which is the bundle of Hilbert spaces with $k$-dimensional fiber $H_{\tilde x}$.
The gauge field $\tilde{A}$ on the dual torus is constructed using the same projection, 
through the Grassmannian connection of the bundle $\tilde E$:
\begin{eqnarray}
P_{\tilde x}\tilde dP_{\tilde x}\chi&=&
    \sum_{p,q} \psi^p\braket{\psi^p|\tilde d(|\psi^q}\braket{\psi^q|\chi})\cr
  &=& \Psi\left[\tilde d\braket{\Psi|\chi}+
        \braket{\Psi|\tilde d\Psi}\braket{\Psi|\chi}\right]~, 
\end{eqnarray} 
where $\tilde d$ is the exterior derivative on the dual torus, which acts on the 
$\tilde{x}$-dependence of the zeromodes.
Thus, the connection of the bundle $\tilde{E}$ is given by
\begin{eqnarray}
  \tilde{A}&=&i\braket{\Psi|\tilde d\Psi},\quad
  \tilde{A}^{pq}_\mu(\tilde{x})=
  i\int_{T^2}\!d^2x\,\psi^{p\dagger}\tilde{\partial}_\mu\psi^q .\label{Nahmtf}
\end{eqnarray} 

From the Atiyah-Singer index theorem \cite{AS}, it is known that there is 
a relation 
\begin{eqnarray}
	{\rm ch}({\rm ind} (D^+_{\tilde{x}}, {\cal E}))=
      \int_{M}\!{\hat{A}}(M)\wedge {\rm ch}({\cal E}).	\label{AS}
\end{eqnarray}
between the Chern character of a bundle $\CE$ over an even spin manifold $M$ with a family of connections carrying a parameter $\tilde x$ ($\tilde x$ being an element of a compact parameter space $Y$),  and  the Chern character of its index bundle. 
$\hat{A}(M)$ is the $A$-roof genus of $M$.
This formula holds for Dirac operators $D^+_{\tilde{x}}: \Gamma(S^+)\rightarrow \Gamma(S^-)$ of any even dimensional spin manifold $M$ coupled to any connection carried by a bundle ${\cal E}$ over $M\times Y$. 
For our case, $M$ corresponds to $T^2$,  ${\cal E}=\pi^* E\otimes {\cal P}$ and the space $Y$  is the dual torus, so that  $\hat{A}(T^2)=1$. Note that the bundle ${\rm ind} (D_{\tilde{x}}, {\cal E})$ on the l.h.s. is nothing but the bundle $\tilde E$ over the dual torus $\tilde{T}^2$. Then eq. (\ref{AS}) boils down to the expression \cite{Hori}
\begin{eqnarray}
	{\rm ch}(\tilde{E})=
      \int_{T^{2}}\!{\rm ch}({\cal P})\wedge {\rm ch}(E)~,	\label{ASind}
\end{eqnarray}
with the Chern characters being defined as
\be
{\rm ch}(E)={\rm Tr}\exp\left(\frac{F}{2\pi}\right)~,~~
{\rm ch}(\CP)=\exp\left({P\over 2\pi}\right)~.	
\ee
Here $F$ is the curvature of the gauge field $A$ and $P$ is the curvature of the Poincar\'e bundle given by $P=\sum_\mu {1\over 2\pi}dx^\mu\wedge d\tilde x_\mu$.
We have set the length of each edge
of the torus and of the dual torus to $2\pi R_i$ and $2\pi \tilde R_i$, respectively, such that$({\rm vol}\,T^{2})\cdot({\rm vol}\,\tilde{T}^{2})=(2\pi)^{4}$.
Thus, the index theorem relates the curvature on the bundle $E$ with 
bundle connection parametrized by the coordinate $\tilde x$ of the dual torus to the bundle $\tilde E$ over the dual torus.

Expanding (\ref{ASind}), we obtain (by considering both sides separately)
\begin{eqnarray}
  (\mbox{L.H.S.}) & = &
       {\rm rank}(\tilde{E})+c_1(\tilde{E}),\nonumber\\
  (\mbox{R.H.S.}) & = &C_1(E)- {\rm rank}(E){d\tilde{x}^1\wedge d\tilde{x}^2\over vol({\tilde T^2})},
\end{eqnarray}
where $c_1(E)$ is the first Chern class and $C_1(E)=\int c_1(E)$ is the Chern number.
From these relations we get
\begin{eqnarray}
  {\rm rank}(\tilde{E}) & = & C_1(E)~,\nonumber\\
  C_1(\tilde{E}) & = & - {\rm rank}(E)~.
\end{eqnarray}
Thus, the index theorem tells us that the Nahm transformation exchanges the rank of the gauge group and first Chern number up to a sign. 
Symbolically, it gives a map of gauge fields of $(N,k)\,\rightarrow\,(k,-N)$.

The minus sign in $-N$ simply reflects the direction of the magnetic field on the dual torus.
From the point of view of the D-brane picture, however, 
it seems naively the appearance of $N$ anti-$D0$-branes instead of $N$ $D0$-branes 
in the T-dual theory.
This discrepancy motivates us to study the consistency with T-duality, which transforms 
$N D2/k D0$-branes into $k D2/N D0$-branes%
\footnote{
In the case of an negative Chern number, $C_1(E)=-k\, (k>0)$, there are $k$-zeromodes 
in the left-handed spinor, and accordingly, the index theorem (\ref{ASind}) is modified 
by multiplying the minus sign in the right hand side, 
so that we have a map $(N,-k)\,\rightarrow\,(k,N)$.
Thus, the rank of the dual gauge group is always kept positive.
Of course, there is the same problem in the D-brane interpretation of the relative sign.
}.

To compare with the T-duality in string theory below, 
we need an explicit profile for the $U(N)$ gauge field $A$ of the constant curvature
and its Nahm transform.
For this end it is convenient to adopt a twisted bundle construction,
originally considered by 't~Hooft \cite{tHooft} and widely used in matrix models 
(see for example \cite{Taylor}).
Define the torus as the covering space $\real^2$ 
quotient by the action of a period lattice $\Lambda =\{ (2\pi R_1 n, 2\pi R_2 m)| n,m\in\integer^2\}$, and decompose  
$U(N)=(U(1)\times SU(N))/{\mathbb Z}_N$.
Then, the gauge field $A$ must obey the boundary condition
\begin{eqnarray}
&&	A_\mu (x_1+2\pi R_1,x_2)
	  =\Omega_1 (x_2)(i\partial_\mu +A_\mu) (x_1,x_2)
       \Omega_1^{-1}(x_2)~,						\nonumber\\
&&  A_\mu (x_1,x_2+2\pi R_2)
      =\Omega_2 (x_1)(i\partial_\mu +A_\mu)(x_1,x_2)
      		\Omega_2^{-1}(x_1)~,		\label{Atf}
\end{eqnarray}
where $\Omega_{\mu}$, $\mu = 1,2$ are $U(N)$-transition functions satisfying the cocycle condition
\begin{eqnarray}
	\Omega_1(2\pi R_2)\Omega_2(0)
    \Omega_1^{-1}(0)\Omega_2^{-1}(2\pi R_1)
    ={\mathbf 1}~.		\label{cocycle}
\end{eqnarray}
The transition functions for the configuration with Chern number $k$ 
can be chosen as
\be
\Omega_1 = e^{ikx_2/NR_2} U^k~,~~~~~\Omega_2=V~,~~~~~
    U^kV=e^{-2\pi ik/N}VU^k~,	\label{gauconfiga}\\
\ee
where the $SU(N)$ matrices $U$ and $V$ are 
\begin{eqnarray}
U=\left(
  \begin{array}{cccc}
    1&&&\\
    &e^{\frac{2\pi i}{N}}&&\\
    &&\ddots&\\
    &&&e^{\frac{2\pi i(N-1)}{N}}\\
  \end{array}
  \right),~~
  V=\left(
    \begin{array}{cccc}
      0&1&&\\
      \vdots&&\ddots&\\
      0&&&1\\
      1&0&\cdots&0
    \end{array}
  \right). \label{matricesUV}
\end{eqnarray}
With this choice of transition functions, (\ref{Atf}) is satisfied 
by the following gauge field configuration
\be
A_1=0~,~~A_2={k\over N2\pi R_1R_2}x_1~, \label{gaugeconfiguration}
\ee
and the field strength is non-vanishing in the $U(1)$ part only, to give the 
desired Chern number:
\be
F_{12}={k\over N2\pi R_1R_2}~,\quad C_1=\frac{1}{2\pi}\int_{T^2} {\rm Tr} F=k~.
\ee
Unlike the 4 dimensional case, the gauge field $A_{\mu}$ has no moduli and 
hence is unique up to $U(N)$ gauge transformations.

With this gauge profile, we now follow the procedure of the Nahm transformation.
The (right-handed) spinors $\psi$ in the $U(N)$ fundamental representation should obey 
the boundary conditions
\begin{eqnarray}
	\psi(x_1+2\pi R_1 , x_2) & = & \Omega_1(x_2) \psi(x_1 , x_2)~,	\nonumber\\
	\psi(x_1 , x_2+2\pi R_2) & = & \Omega_2(x_1) \psi(x_1 , x_2)~,	\label{mattrf}
\end{eqnarray}
and the general solution of these conditions is given by \cite{Taylor}.
 \bea
  \psi(x_1,x_2)
&=&\sum_{s\in{\mathbb Z}}\sum^k_{p=1}
  \exp\left[
    i{x_1\over R_1}
      \frac{k}{N}Y
  \right]
  \phi^p\left(Y
  \right),	\label{psi}
\eea
where $Y \equiv \frac{x_2}{2\pi R_2} + u + Ns + \frac{N}{k}p$, and 
$\phi^p \,(p=1,\cdots,k)$ is an arbitrary function of $Y$.
Here $u= 1,\cdots,N$ labels each component of $SU(N)$ fundamental representation.
By substituting (\ref{psi}) into the Dirac equation (\ref{Dirac}), 
it reduces to equations for $\phi^p$. 
For a fixed $p$ and $u$ (they are implicitly inside $Y$) it is
\begin{eqnarray}
  \left(
    \frac{k}{NR_1}Y - {\tilde{x}_1 + i\tilde{x}_2\over2\pi} + \frac{1}{2\pi R_2}\partial_Y
  \right)
  \phi^p(Y) = 0~,
\end{eqnarray}
which is solved as 
\begin{eqnarray}
  \phi^p(Y)
  =f^p(\tilde{x}_1,\tilde{x}_2)
    \exp\left[
      -\frac{\pi kR_2}{NR_1}Y^2 + R_2 \left( \tilde{x}_1 + i\tilde{x}_2 \right) Y
    \right],
\end{eqnarray}
where $f^p$ is an arbitrary function. 
For generic $f^p$, we get $k$ normalized zeromodes $\xi^p$ ($p=1,\cdots,k$), 
whose component $\xi^p_u$ is given by
\be
  \xi^p_u(x,\tilde{x})
  =\CN\sum_{s\in {\mathbb Z}}
    \exp\left[
      ix_1\frac{k}{NR_1}Y 
   + R_2 i\tilde{x}_2Y 
    -\frac{\pi kR_2}{NR_1}\left\{
     Y -\frac{NR_1}{2\pi k}\tilde{x}_1
    \right\}^2
  \right],
\ee
where $\CN= \left( \frac{k}{8\pi ^4NR_1^3R_2} \right)^{\frac{1}{4}}$
is the normalization constant to satisfy the orthonormal condition 
$\sum_u \int d^2 x \xi_u^{p\dag} \xi_u^q = \delta^{pq}$.
The general solution for the Dirac zero modes $\psi^p$ is a linear combination of these
of the form
\be
  \psi^p (x,\tilde{x})=\sum_q \xi^q (x,\tilde{x}) g^{qp} (\tilde{x})~,
\ee
where $g$ is a $U(k)$-valued function on $\tilde{T}^2$, which originates from $f^p$ above.
 
Substituting this result into formula (\ref{Nahmtf}), 
we obtain the $U(k)$ gauge fields as $k\times k$ hermitian matrix valued fields as
\begin{eqnarray}
  \tilde{A}_1 & = & - (g^\dagger \tilde{\partial}_1 g)(\tilde{x}_1,\tilde{x}_2)~,\nonumber\\
  \tilde{A}_2 & = & -\frac{ NR_1R_2}{2\pi k}\tilde{x}_1 {\bf 1}_k  
  				  - (g^\dagger \tilde{\partial}_2 g)(\tilde{x}_1,\tilde{x}_2).
\end{eqnarray}
By using the $U(k)$ gauge freedom
we can put $g(\tilde x_1,\tilde x_2)=1$.
Then, the gauge configuration on the dual torus  
$\tilde{T^2}$ is
\begin{eqnarray}
	\tilde{A}_1=0~,~~~
    \tilde{A}_2 = - \frac{N}{2\pi\tilde R_1\tilde R_2k} \tilde{x}_1~,~~~
    \tilde{F}_{12} = -\frac{ N}{2\pi\tilde R_1\tilde R_2 k}~,~~~\tilde{C}_1 = -N.
    \label{dual gauge config}
\end{eqnarray}
From this we can read off the transition functions as 
\begin{eqnarray}
  \tilde{\Omega}_1=e^{- iN \tilde{x}_2/k\tilde{R}_2}\tilde{U}^N~,~~~
  \tilde{\Omega}_2=\tilde{V}~,
  \label{dual transition}
\end{eqnarray}
where $\tilde{U}$ and $\tilde{V}$ are $SU(k)$ matrices similar to (\ref{matricesUV}).
This is the configuration of a gauge group $U(k)$ with
first Chern number equal to $-N$, 
in accordance with the
Atiyah-Singer family index theorem. 
We have also shown the correspondence of two moduli spaces of constant curvature 
connections on the torus and the dual torus.

We close this section with a historical remark.
To our knowledge, \cite{Kaji} is the first paper that
teated the $2$-dimensional version of the Nahm transformation concretely.
Our presentation here follow closely that paper, 
but we correct their intermediate calculations, 
leading to the important sign in $\tilde{C}_1=-N$.

\section{The boundary state on the torus $T^2$}
\label{sec:boundary}

In this section we 
consider boundary states of $D2/D0$ bound states on the torus
and investigate the compatibility between T-duality and the Nahm transformation.
We first follow the construction of the boundary state of bosonic strings 
developed in \cite{Duo,DiVecc}, and take its T-dual. 
It turns out that they are consistent in the bosonic string sector.
However, this raises a puzzle for the interpretation of the result from the
D-brane point of view.
We thus generalize the argument to superstring theory 
by focusing on the RR-sector.
To this end, we give a novel representation of the RR zeromodes, that is suitable to 
deal with T-duality.
We will show the invariance of the Chern-Simons term for the bound state 
and resolve the puzzle at the end of this section.

Our parametrization here is such that the two edges of the torus have lengths  $2\pi \sqrt{\alpha'}$ . Then the information of the shape of the torus which appeared in the previously discussed Nahm transformation is now absorbed into the metric.

\subsection{Construction of the boundary state}

It is known that in Minkowski space, 
the boundary state $\Ket{B}$ for a single $Dp$-brane is given by 
a sum of eigenstates for all loops on the worldvolume \cite{Callan}. 
A boundary state describing a system of $Dq$-branes inside $Dp$-branes 
is constructed by operating an appropriate Wilson loop factor $\CO_A$, carrying the 
information on the $Dq$-branes as a non-trivial gauge connection,  
on the boundary state of a single $Dp$-brane.
In \cite{Duo,DiVecc}, this construction is generalized to 
the case of a toroidal target space $T^{2d}$ and a non-trivial gauge bundle 
wrapping a torus,
called a magnetized $D(2d)$-brane.
Note that a string loop around a cycle of the torus is
described by an open path in the corresponding covering space.
In such a case, a naive Wilson line factor is not gauge invariant.
To obtain a gauge invariant loop factor $\CO_A$, 
transition functions of the gauge bundle have to be inserted along 
this path in the covering space
\cite{Duo,DiVecc}.

More specifically, 
the boundary state in the direction of a magnetized $D2$-brane wrapping 
$N$ times a torus $T^2$ is given by
\bea
\left|B_{F}\right>&=&{\cal O}_A\left|B\right>
\cr
&=& {\cal O}_A\sqrt{\det( G + B )}\sum_{\vec m\in{\mathbb Z}^2}^{}\prod^{\infty}_{n=1}
    e^{-\frac{1}{n}\alpha_n^\dag GR\tilde{\alpha}^\dag_n}
    \left|\vec 0; \vec m\right>,\label{finalanswer}
\eea
Here $\Ket{B}$ is the standard Neumann boundary state on $T^{2d}$,
where $\Ket{\vec n;\vec m}$ denotes a state for zeromodes 
with the Kaluza-Klein momentum $\vec n\in{\mathbb Z}^2$ and 
the winding numbers $\vec m\in{\mathbb Z}^2$ along $T^2$.
$G_{\alpha\beta}$ and $B_{\alpha\beta}$ are the metric and the $B$-field on $T^2$, 
respectively, and their combination $R= (G+B)^{-1}(G-B)$ indicates the boundary 
condition $((G+B)\alpha_n +(G-B)\tilde{\alpha}_{-n})\Ket{B} =0$.
Next, the Wilson loop factor $\CO_A$ is given by \cite{Duo}
\be
  {\cal O}_A = {\rm Tr}_N \prod_{\alpha =1,2}\prod^{{\mathbf m}^{\alpha}-1}_{\ell=0}
    \Omega_\alpha ({\mathbf x}+2\pi \sqrt{\alpha'}\sum^{\alpha-1}_{\beta=1}
{\mathbf m}^\beta a_\beta +2\pi \sqrt{\alpha'}\ell a_\alpha)
  \exp(-S_A).            \label{modiwil}
\ee
where the exponential factor is a naive path-ordered Wilson loop operator specified by 
a $U(N)$ gauge configuration
\be
\exp(-S_A) = P \exp\left({i\over 2\pi{\alpha}'}\int_0^{2\pi} 
A_\alpha \partial_{\sigma}X^\alpha d\sigma\right)~,
\ee
and $\Omega_\alpha (x)$ is a transition function of the gauge bundle.
In the above, $X^\alpha(\sigma)$ (string variable), 
${\mathbf x}^\alpha$ (center of mass coordinate) and ${\mathbf m}^\alpha$ (winding)
are operators acting on the closed string Hilbert space.
To understand the expression, let us act ${\cal O}_A$ on a eigenstate of $X(\sigma)$, 
which is seen as a path in the covering space $\real^2$ starting from 
$x+2\pi \sqrt{\alpha'}\sum_{\alpha=1}^2 m^\alpha a_\alpha$ ($m_\alpha \in \integer$) 
and ending in $x$,
with $a_\alpha$ being the $\alpha$-th cycle of the torus $T^2$.
Whenever the path passes across the boundary of a cell, 
a transition function must be introduced to glue the fields.
The above factors correctly take this effect into account to maintain 
the Wilson loop factor gauge invariant. 
Because the Neumann boundary state is a sum over all loops, 
eigenvalues $x^\alpha$ and $m^\alpha$ are replaced with corresponding operators.
An equivalent and more direct definition is also given in \cite{DiVecc}.


The significance of this construction is already seen in the case of $N D2$-branes 
with vanishing gauge field $A_\alpha=0$.
Consider a $D2$-brane wrapped $N$ times around the cycle $a_2$ along $x^2$-direction.
In this case, transition functions can be chosen as
$\Omega_1 =1, \Omega_2 = V$, and the Wilson loop factor  
\be
{\cal O}_{A=0} = {\rm Tr}_N (V^{{\mathbf m}_2})
\ee
constrains possible winding numbers along $x^2$ to $m_2 =Ns_2 $ ($s_2 \in \integer$)
integer multiples of the wrapping number $N$.
Then, the corresponding boundary state is obtained as
\begin{eqnarray}
  \left| B_{A=0} \right>&=&
   N \sqrt{\det( G + B )}\sum_{\vec s\in{\mathbb Z}^2}^{}\prod^{\infty}_{n=1}
    e^{-\frac{1}{n}\alpha_n^\dag GR\tilde{\alpha}^\dag_n}
    \left|0;
	\left(
      \begin{array}{c}
         s_1\\Ns_2
      \end{array}\right)
	\right>. \label{A0bound}
\end{eqnarray}
Since the $D2$-brane is wrapped $N$ times along $x^2$-direction, 
closed strings emitted from it carry winding numbers along this direction 
with integer multiples of $N$.
Suppose now, a $D2$-brane wrapped $N_\alpha$ times along the $\alpha$-th cycle 
of the torus with $N=N_1 N_2$. 
Then, the transition functions are 
$\Omega_1=V_{N_1\times N_1}\otimes 1_{N_2\times N_2}$ and 
$\Omega_1=1_{N_1\times N_1}\otimes V_{N_2\times N_2}$, and thus by acting 
\be
{\cal O}_{A=0} = {\rm Tr}_N (\Omega_1^{{\mathbf m}_1}\Omega_2^{{\mathbf m}_2}),
\ee
we obtain
\begin{eqnarray}
  \left| B_{A=0} \right>&=&
    N \sqrt{\det( G + B )}\sum_{\vec s\in{\mathbb Z}^2}^{}\prod^{\infty}_{n=1}
    e^{-\frac{1}{n}\alpha_n^\dag GR\tilde{\alpha}^\dag_n}
    \left|0;
\left(
      \begin{array}{c}
         N_1s_1\\N_2s_2
      \end{array}\right)
\right>. \label{A0bound2}
\end{eqnarray}
These examples shows the importance of transition functions to determine the zeromode
structure in the boundary state.

\subsection{T-duality transformation - bosonic sector}

Let us construct the boundary state on $T^2$ corresponding to the situation of 
the Nahm transformation in the previous section.
The $N D2/k D0$ bound state is regarded as a $D2$-brane wrapped $N$ times on $T^2$, 
with a gauge bundle with first Chern number $k$ on it. Here, we only consider 
the case where $N$ and $k$ are coprime.
This is specified by the $U(N)$ gauge configuration (\ref{gaugeconfiguration}) 
\begin{eqnarray}
&&  A_1=0~,~~A_2 = \frac{k}{2\pi\alpha' N}x_1~,~~~~
    F_{12}=\frac{k}{2\pi\alpha' N}~,~~~C_1=k,
    \label{U(N)gauge config}
\end{eqnarray}
as well as the transition functions (\ref{gauconfiga}) 
\be
\Omega_1 = e^{ikx_2/N\sqrt{\alpha'}} U^k~,~~~~~\Omega_2=V~.\label{transition}
\ee
Note that the convention is slightly changed, 
with the data about the shape of the torus being encoded in the parameters $a_i$ of the metric, which give the length of the $i$-cycle in the unit of $\sqrt{\alpha'}$.
Thus we take flat background as 
\be
G=\left(
\begin{array}{cc}
	a_1^2&0\\
	0&a_2^2
\end{array}
\right), ~~~
a_i=\frac{R_i}{\sqrt{\alpha'}}
\ee 
and $B=0$.

The boundary state for this kind of situations is already computed in \cite{Duo,DiVecc}, so that we just quote the result.
It is equivalent to replace $B$ in the Neumann boundary state with $2\pi \alpha^\prime F$:
\begin{eqnarray}
  \left| B_{F} \right> &=&
   N\sqrt{{\rm det}(G+2\pi\alpha^\prime F)}
    \sum^{}_{\vec s\in{\mathbb Z}^{2}}
    e^{-i \pi {\mathbf m}^1 (2\pi \alpha^\prime F_{12}){\mathbf m}^2}
\nonumber\\
&&\hspace{3em}   \times 
       \prod^{\infty}_{n=1}
       e^{-\frac{1}{n}\alpha_n^\dag 
         G{\cal R}
         \tilde{\alpha}_n^\dag}
  \left|-(2\pi \alpha' F)N\vec s,N\vec s\,\right>~,
  \label{B_F general}
\end{eqnarray}
where ${\cal R}=(G+2\pi\alpha^\prime F)^{-1}(G-2\pi\alpha^\prime F)$ represents 
the mixed boundary condition.
Note that all $F$'s in this expression belong to the $U(1)$ part of the $U(N)$ field strength.
The zeromode part in components is $(n_1, n_2)=(-2\pi\alpha'F_{12}Ns_2,2\pi\alpha'F_{12}Ns_1)$ and 
$(m_1, m_2)=(Ns_1,Ns_2)$, labeled by two integers $s_{1,2}$, respectively.
This structure of the zeromodes and the extra phase factor in the first line above
are the consequence of modifying the Wilson loop operator.
The result depends, however, only on the metric $G$, the rank $N$ and the $U(1)$ part of the field strength $F_{12}$.

By substituting the explicit form of $G$ and $F$ into (\ref{B_F general}), we obtain
\begin{eqnarray}
  \left| B_{F} \right> 
  &=&\sqrt{(a_1 a_2 N)^2+k^2}
  \sum^{}_{\vec s\in{\mathbb Z}^{2}}
    e^{-i\pi Nks_1s_2}
   \prod^{\infty}_{n=1}
       e^{-\frac{1}{n}\alpha_n^\dag 
         G{\cal R}
         \tilde{\alpha}_n^\dag}
   \left|\left(
      \begin{array}{c}
         -ks_2\\ks_1
      \end{array}\right),\left(
      \begin{array}{c}
        Ns_1\\Ns_2\\
      \end{array}\right)\right>,\nonumber\\
G{\cal R}&=&
   \displaystyle{\frac{1}{a_1^2 a_2^2 N^2+k^2}} 
        \left(
          \begin{array}{cc}
           a_1^2 & 0\\ 0 &a_2^2
          \end{array}\right)
        \left(
          \begin{array}{cc}
           a_1^2 a_2^2 N^2- k^2
             &-2a_2^2 Nk\\
           2a_1^2 Nk
             &a_1^2 a_2^2 N^2- k^2
          \end{array}\right).
     \label{boundstate}
\end{eqnarray}
This is the boundary state for $N D2/k D0$ bound state on $T^2$.
Except for the phase factor, this form is also determined by solving 
the mixed boundary condition.  
The square-root factor is related through $\braket{{0;0}|B_F}$ to 
the effective Dirac-Born-Infeld action, 
and it is a typical mass formula for the BPS bound state.

We now take the T-duality transformation along both directions of $T^2$. 
The radii of the dual torus $\tilde{T}^2$ are encoded in the dual metric 
\be
\tilde{G}=\left(
\begin{array}{cc}
	\tilde{a}_1^2&0\\
	0&\tilde{a}_2^2
\end{array}
\right)
=\left(
\begin{array}{cc}
	1/a_1^2&0\\
	0&1/a_2^2
\end{array}
\right) =G^{-1},
\ee 
which is nothing but the Buscher rule for the case of a diagonal metric $G$ 
and $B=0$.
T-duality also acts on oscillators as%
\footnote{
Here the arrow means the following procedure:
first rewriting an original operator $\alpha_n (G)$ defined in a background
metric $G$ in terms of dual operators as $\alpha (G)=\tilde{G}\alpha'_n (\tilde{G})$, 
then omitting the prime for notational simplicity.
The invariance of the spectrum under the T-duality is written 
in this notation as $H\sim \alpha^t_{-n}G\alpha_n \rightarrow 
\alpha^t_{-n}\tilde{G}\alpha_n$, i.e., the closed-string Hamiltonian is 
written in two different ways.
}
\be
(\alpha_n, \tilde{\alpha}_n)
\to (\tilde{G}\alpha_n, -\tilde{G}\tilde{\alpha}_n).
\label{oscilltor dual}
\ee
This, in particular, exchanges momenta and winding operators 
$({\mathbf n},{\mathbf m}) \to ({\mathbf n},{\mathbf m})$,
as well as their eigenstates 
\be
\Ket{\vec n;\vec m}\to e^{i\pi {\mathbf m}^t{\mathbf n}}\Ket{\vec m;\vec n},
\label{zeromode dual}
\ee
Note the extra phase factor, first proposed in \cite{Duo}.
It is needed for the T-duality invariance of the cocycle factor 
associated with the 3-string vertex.

Under this transformation, the boundary state (\ref{boundstate}) is rewritten as follows.
First by using (\ref{oscilltor dual}),
 the oscillator part in the exponent becomes 
$\alpha^t_n G{\cal R} \alpha_{-n}\to \alpha^t_n \tilde{G}^t G {\cal R} (-\tilde{G} \alpha_{-n})
=\alpha^t_n (-{\cal R} \tilde{G}) \alpha_{-n}$, and $-{\cal R} \tilde{G}$ is written by dual 
variables as
\begin{align}
-{\cal R} \tilde{G}
&=\frac{-1}{N^2/\tilde{a}_1^2 \tilde{a}_2^2 +k^2}
        \left(
          \begin{array}{cc}
          N^2/\tilde{a}_1^2 \tilde{a}_2^2- k^2
             &-2Nk/\tilde{a}_2^2 \\
           2Nk /\tilde{a}_1^2 
             &N^2/\tilde{a}_1^2 \tilde{a}_2^2- k^2
          \end{array}\right)
        \left(
          \begin{array}{cc}
           \tilde{a}_1^2 & 0\\ 0 &\tilde{a}_2^2
          \end{array}\right)\nonumber\\
&=\frac{1}{\tilde{a}_1^2 \tilde{a}_2^2 k^2+N^2}
        \left(
          \begin{array}{cc}
          \tilde{a}_1^2 \tilde{a}_2^2 k^2-N^2
             &2Nk\tilde{a}_1^2 \\
           -2Nk\tilde{a}_2^2 
             &\tilde{a}_1^2 \tilde{a}_2^2 k^2-N^2
          \end{array}\right)
        \left(
          \begin{array}{cc}
           \tilde{a}_1^2 & 0\\ 0 &\tilde{a}_2^2
          \end{array}\right)\nonumber\\ 
&=\frac{1}{\tilde{a}_1^2 \tilde{a}_2^2 k^2+N^2}
        \left(
          \begin{array}{cc}
           \tilde{a}_1^2 & 0\\ 0 &\tilde{a}_2^2
          \end{array}\right)
        \left(
          \begin{array}{cc}
          \tilde{a}_1^2 \tilde{a}_2^2 k^2-N^2
             &2Nk\tilde{a}_2^2 \\
           -2Nk\tilde{a}_1^2 
             &\tilde{a}_1^2 \tilde{a}_2^2 k^2-N^2
          \end{array}\right). 
\end{align}
The square-root factor is rewritten similarly, and the phase factor is invariant.
Next, according to (\ref{zeromode dual}), a zeromode state 
in (\ref{boundstate}) becomes 
\be
\left|\left(
      \begin{array}{c}
         -ks_2\\ks_1
      \end{array}\right),\left(
      \begin{array}{c}
        Ns_1\\Ns_2\\
      \end{array}\right)\right>
\to \left|\left(
      \begin{array}{c}
        Ns_1\\Ns_2\\
      \end{array}\right),\left(
      \begin{array}{c}
         -ks_2\\ks_1
      \end{array}\right)\right>,
\ee
that is, the cocycle factor is trivial in this case.
Finally, by renaming $t_1=-s_2$ and $t_2=s_1$, we arrive at
\begin{eqnarray}
  \left| \widetilde{B_{F}} \right> 
  &=&\sqrt{(\tilde{a}_1 \tilde{a}_2 k)^2+N^2}
  \sum^{}_{\vec t\in{\mathbb Z}^{2}}
    e^{i\pi Nkt_1t_2}
   \prod^{\infty}_{n=1}
       e^{-\frac{1}{n}\alpha_n^\dag 
         \tilde{G}\tilde{{\cal R}}
         \tilde{\alpha}_n^\dag}
   \left|\left(
      \begin{array}{c}
         Nt_2\\-Nt_1
      \end{array}\right),\left(
      \begin{array}{c}
        kt_1\\kt_2\\
      \end{array}\right)\right>,\nonumber\\
\tilde{G}\tilde{{\cal R}}&=&
   \displaystyle{\frac{1}{\tilde{a}_1^2 \tilde{a}_2^2 k^2+N^2}} 
        \left(
          \begin{array}{cc}
           \tilde{a}_1^2 & 0\\ 0 &\tilde{a}_2^2
          \end{array}\right)
        \left(
          \begin{array}{cc}
           \tilde{a}_1^2 \tilde{a}_2^2 k^2- N^2
             &2\tilde{a}_2^2 kN\\
           -2\tilde{a}_1^2 kN
             &\tilde{a}_1^2 \tilde{a}_2^2 k^2- N^2
          \end{array}\right).
     \label{T2dual}
\end{eqnarray}

By comparing this with (\ref{boundstate}), we see that the T-duality is 
the replacement $(N,k) \to (k,-N)$ of two parameters%
\footnote{
Another replacement $(N,k)\to (-k,N)$ is possible, but is equivalent to the above 
by a further renaming of $t$'s.
We have chosen $(N,k) \to (k,-N)$ so as to maintain the rank of the 
dual gauge group to be positive.
On the other hand, if $C_1(E)=-k$, this kind of renaming is the right one.
See the remark in the footnote in \S\ref{sec:Nahm}.
}, 
in addition to the change of background
$a_\alpha \to \tilde{a}_\alpha$.
Moreover, the same boundary state (\ref{T2dual}) on the dual torus 
is obtained by starting from Nahm-transformed gauge configuration 
(\ref{dual gauge config}), (\ref{dual transition}) with gauge group $U(k)$ and the 1st Chern number $-N$:
\begin{eqnarray}
	\tilde{A}_1=0~,~~~
    \tilde{A}_2 = - \frac{N}{2\pi \alpha' k} \tilde{x}_1~,~~~
    \tilde{F}_{12} = -\frac{ N}{2\pi\alpha' k}~,~~~\tilde{C}_1 = -N,
\end{eqnarray}
\begin{eqnarray}
  \tilde{\Omega}_1=e^{- iN \tilde{x}_2/k\sqrt{\alpha'}}\tilde{U}^N~,~~~
  \tilde{\Omega}_2=\tilde{V}~,
\end{eqnarray}
with suitable redefinitions.
This is evident by following the same procedure leading to (\ref{boundstate}).
Therefore, this result shows the consistency between T-duality and 
the Nahm transformation, importantly including its sign. 

From the D-brane point of view, however, this sign seems problematic,
since the $D0$-brane charge naively equals to $C_1(\tilde{E})=-N$,
that suggests $N$ anti-$D0$-branes.
To see the correct $D0$-brane charge, we should work with superstring theory,
and investigate the T-duality rule for RR-potentials to be careful with its sign.

\section{Fermionic sector and T-duality}

So far we found a complete agreement of the Nahm transformation and the T-duality transformation  in the bosonic sector of the boundary state.
Here, we extend this analysis to the fermionic part of the boundary state:
To construct the fermionic boundary state for $ND2/kD0$-branes on $T^2$ and 
to perform the T-duality transformation.

The fermionic counterpart of a boundary state (\ref{finalanswer})
with and without gauge flux has been already formulated in refs.
\cite{Callan,PolCai88,Billo,9707068}.
There, the possible effects of transition functions in (\ref{modiwil}) are not 
taken into account.
But it turns out that there is no such effect, and their result is still correct.
There, the NSNS-oscillators $(\psi_r, \tilde{\psi}_r)$ ($r \in \integer +1/2$) and the non-zero RR-oscillators $(\psi_n, \tilde{\psi}_n)$ ($n \ne 0 $) 
contribute to the boundary state in completely the same way as the bosonic sector, 
written by the matrix ${\cal R}$ as in (\ref{boundstate}).
The T-duality transformation of this part is also the same as the bosonic case.
Therefore, the agreement of the Nahm transformation and the T-duality transformation
in these parts is clear.

According to the remark above, in this section, 
we focus on the RR-zeromode part of the boundary state.
and the coupling of the D-branes to the RR antisymmetric fields.
We will first give the boundary state for $D2/D0$-branes, 
by using a novel representation of RR-vacuum, alternative to
the standard formulation in \cite{Billo,9707068}. 
Then we describe its T-duality transformation.
After the introduction of the conjugate state in a new representation, 
we study the RR-coupling of the D-branes.
The T-duality invariance of the Chern-Simons term is shown in a remarkably 
simple form.
In the following, the shape of the torus is less important, 
so that we simply set the metric $G$ as $a_1=a_2=1$
(see also the remark below on this point).

\subsection{Boundary state for RR zeromodes}

We first describe a new representation for the RR zeromodes briefly.
See the appendix for more details.
Then it is applied to construct D-brane boundary states.

As is well-known, 
the world sheet fermion 
($\psi^\mu$ for left mover and 
$\tilde\psi^\mu$ for right mover) in the R-sector has mode expansion labeled by integers,
and its zeromodes satisfy the $10$ dimensional Clifford algebra. 
Accordingly, 
the ground states of the left (right) movers are specified 
by spinor indices $\Ket{A}$($\Ket{\tilde B}$) of $SO(1,9)$, respectively, 
where $A,B$ run $1,\cdots,32$. 
A RR-ground state is a linear combination of 
tensor products $\Ket{A}\otimes\Ket{\tilde B}$.
In the present paper, an explicit gamma matrix representation is taken as
\bea
\psi_0^\mu\Ket{A}\otimes\Ket{\tilde B}
&=&({1\over\sqrt{2}}\Gamma^\mu\otimes\id
)\Ket{A}\otimes\Ket{\tilde B}~,\cr
\tilde\psi_0^\mu\Ket{A}\otimes\Ket{\tilde B}
&=&(\Gamma_{11}\otimes{i\over\sqrt{2}}\Gamma_{11}\Gamma^\mu
)\Ket{A}\otimes\Ket{\tilde B}~.
\eea
Note an extra $i \Gamma_{11}$ in the right movers,
as compared to the standard representation \cite{Billo,9707068} (see also \cite{KW}.).
This is merely one choice of a representation of 
the $SO(1,9)\times SO(1,9)$ Clifford algebra, 
but its advantages will become apparent when we take the GSO projection into 
account (see also the appendix).

Define an operator $\theta^\mu$ as
\be
\theta^\mu=(\psi^\mu_0+i\tilde\psi^\mu_0)/\sqrt{2},\label{thetaoperators}
\ee
satisfying the anti-commutation relation 
\be
\{\theta^\mu,\theta^\dagger_\nu\}=\delta^\mu_\nu,
\ee
and define the vacuum state
$\Ket{[C]}$ in the asymmetric picture of left and right movers \cite{Yost} as
\be
\Ket{[C]}\equiv e^{i\beta_0\tilde\gamma_0}
\Ket{A,-1/2}C_{AB}\Ket{\tilde B,-3/2}~,
\ee
where $C_{AB}$ is the charge conjugation matrix.
Here we added the bosonic ghost sector in the superstring, 
and $\beta_0$, $\gamma_0$ are their zeromodes. 
We take the picture of the left mover as $-\half$ and of the right mover as $-{3\over2}$ to saturate the background charge of the $\beta\gamma$ ghosts. 
Then, it is easy to show that
\be
\theta^\mu \Ket{[C]}=0 ~,~~\theta^\dagger_\mu \Ket{[C]}=\Ket{[-C\Gamma_\mu]}~.
\ee
Any bispinor state is made by acting creation operators $\theta^\mu$
on the vacuum state $\Ket{[C]}$.

In particular, a state
\be
\Ket{Dp}=\theta_p^\dagger\theta_{p-1}^\dagger\cdots\theta_0^\dagger\Ket{[C]}
\label{Dp}
\ee
satisfies the boundary condition for a $Dp$ brane
\begin{align}
& \theta^\dagger_\alpha\Ket{Dp}=0, ~~(\alpha=0,\cdots,p), \nonumber\\
& \theta_i\Ket{Dp}=0,~~(i=p+1,\cdots,9).
\label{Dp bc}
\end{align}
Therefore, $\Ket{Dp}$ can be identified with the zeromode part of the $Dp$-brane%
\footnote{To consider the GSO projection, we need to consider the two types of boundary conditions which are labeled by $\eta=\pm1$. Here, we only write the formula for the  $\eta=+1$ case, and do not write the $\eta$ explicitly. See the appendix.}.
In particular, the vacuum state $\Ket{[C]}$ is identified with the $D$-instanton.

A gauge field on $N D$-branes 
is introduced through the superstring version of the Wilson loop factor 
in (\ref{finalanswer}).
For a constant flux $F_{\alpha\beta}$, its contribution on RR-zeromodes
is the factor $e^{\pi \alpha' F_{\alpha\beta}\theta^\alpha\theta^\beta}$ 
inside the trace in (\ref{modiwil}),
where $\alpha,\beta$ denote the directions along the $D$-brane.
In our case of $ND2/kD0$-branes wrapping the torus $T^2$, the flux has $U(1)$ part only,
and this factor can be put outside the trace.
Thus, the relevant part of the boundary state for $ND2/kD0$-branes is given by
\bea
\Ket{D2D0}
&=&N e^{2\pi\alpha'F_{12}\theta^1\theta^2}\Ket{D2}
\cr&=&N(1+2\pi\alpha'F_{12}\theta^1\theta^2)\Ket{D2}
\cr&=&N\Ket{D2}+k\Ket{D0}~,
\label{ND2kD0}
\eea
where (\ref{U(N)gauge config}) is inserted.
We included the factor $N$ coming from the trace
which was already present in (\ref{boundstate}).
This state satisfies the required mixed boundary condition
\be
\theta_\alpha^\dagger\Ket{D2D0}=2\pi\alpha'F_{\alpha\beta}\theta^\beta\Ket{D2D0}~,
\ee
where $\alpha,\beta=0,1,2$ and $F_{0\alpha}=0$.

Now we perform the T-duality transformation on this state.
The T-duality of the oscillator part of the fermion is obtained by 
changing the sign of the right mover oscillator mode $\tilde \psi_n \to -\tilde \psi_n$.
On the other hand, T-duality transformation of the zeromode for 
the $\alpha$-direction can be
represented by an operator ${\cal T}_\alpha$ as 
\be
{\cal T}_\alpha =\theta^\alpha -\theta^{\alpha\dagger},
\label{T-dual op}
\ee
which maps ${\cal T}_{\alpha}\theta^\beta{\cal T}^\dagger_\alpha =-\theta^{\dagger\beta}$
and ${\cal T}_{\alpha}\theta^{\dagger\beta}{\cal T}^\dagger_\alpha =-\theta^{\beta}$,
and satisfies ${\cal T}^{\dagger}_{\alpha}{\cal T}_{\alpha}=1$ and 
${\cal T}_{\alpha}{\cal T}_{\alpha}=-1$.
This kind of transformation is used in \cite{Fukuma} for the T-duality rule of RR-potentials, in the context of the chiral spinor representation of RR-potential 
under the T-duality group \cite{Zumino,Hassan}.
One easily checks that the ${\cal T}_\alpha$ maps a boundary state (\ref{Dp}) of 
a $Dp$-brane to that of a $D(p+1)$ or a $D(p-1)$ brane, depending on whether the 
direction of the T-duality map
is perpendicular or parallel to the $Dp$ brane, respectively.

In our case, 
we take the T-duality of (\ref{ND2kD0}) for both directions along the $T^2$.
By applying the T-duality operator ${\cal T}_2{\cal T}_1$, it is easy to check that
\be
\Ket{D2D0}'
={\cal T}_2{\cal T}_1\Ket{D2D0}
=k \Ket{D2}-N \Ket{D0}~.
\label{kD2-ND0}
\ee
This is again the replacement $(N,k)\to (k,-N)$.
Thus, the Nahm transformation of the $T^2$ and the T-duality transformation of the
boundary state are also consistent in the fermionic sector.

Note that there are several sign (or phase) ambiguities
in the above construction.
First, the sign of (\ref{Dp}) in the definition is ambiguous, because it is 
only constrained by boundary conditions (\ref{Dp bc}).
Next, there is a phase ambiguity in the definition of the 
T-duality transformation (\ref{T-dual op}), in general. 
In fact, our choice includes the minus sign in
${\cal T}_{\alpha}\theta^\beta{\cal T}^\dagger_\alpha =-\theta^{\dagger\beta}$.
Finally, our ordering ${\cal T}_2{\cal T}_1$ of the T-duality along 
$T^2$ is minus of another ordering ${\cal T}_1{\cal T}_2$.
All these ambiguities would change sign in boundary states before (\ref{ND2kD0})
and after the T-duality (\ref{kD2-ND0}).
However, the relative sign change encoded in $(N,k)\to (k,-N)$ is always true, 
irrespective of 
the choices discussed above.

\subsection{Consistency of the Chern-Simon coupling}

The RR-charges carried by a boundary state of $D$-branes are 
measured by the coupling to closed-string states of RR-potentials.
Here, we construct such states and study their T-duality transformation,
then we show that the Chern-Simons coupling is T-duality invariant.
After that, we compare it with the existing T-duality rule at the 
low energy effective theory.

As is argued in \cite{Billo} and also in \cite{Berkovits}, 
the RR-state for a RR $q$-form gauge potential
$A=A_{\mu_1\cdots \mu_q}dx^{\mu_1}\cdots dx^{\mu_q}$
 (not field strength)
in the asymmetric picture can be constructed on the vacuum 
state $\Ket{[C]}$. 
To calculate the overlap we need its conjugate state, which we define 
as follows: 
\be
\Bra{[C]}
=\Bra{\tilde A,-1/2}[C]_{AB}\Bra{B,-3/2}\label{bravacuum}e^{-i\beta_0\tilde\gamma_0}~,
\ee
where the action of the fermionic zeromodes on the tensor state are
\be
 \Bra{\tilde A}\otimes\Bra{B} \psi_\mu={-1\over\sqrt{2}}\Bra{\tilde A}
\otimes\Gamma^B_{\mu,C} \Bra{C}~,
\ee
\be
\Bra{\tilde A}\otimes\Bra{B}\tilde \psi_\mu 
={i\over\sqrt{2}}(\Gamma_{11}\Gamma_\mu)^A{}_C\Bra{\tilde C}\otimes\Gamma^B_{11,D}\Bra{ D}~.
\ee
One can see that the state (\ref{bravacuum}) satisfies
\be
\Bra{[C]}\theta_\mu^\dagger=0~,~~\Bra{[C]}\theta^\mu=\Bra{[-C\Gamma^\mu]}~.
\ee
The RR-state for antisymmetric tensor potentials can be defined as
\be
\Bra{\CA}=\Bra{[C]}\CA
=\Bra{[C]}\sum A_{\mu_1\cdots\mu_q}\theta^{\mu_1}\cdots\theta^{\mu_q}~.\label{RRstate}
\ee
At this stage the summation is over any $q$-forms ($q=0,1,\cdots,10$).
To restrict the state to the physical state we perform the GSO projection:
\be
[GSO]={(1-(-1)^{F_0+G_0})(1+(-1)^{p+\tilde F_0+\tilde G_0})\over4}~,\label{GSOproj}
\ee
which also selects the rank of the antisymmetric tensor field
according to the type of the superstring ($p$ even/odd for type IIA/IIB, respectively). 
In general, the state (\ref{RRstate}) depends 
on bosonic zeromodes (momentum/winding) $\Bra{n;m}$,
however, it is sufficient to consider states with $\Bra{0;0}$ in the following.

The T-duality transformation of the RR-state is also given by the operators ${\cal T}_\alpha$ 
(\ref{T-dual op})%
\footnote{
In general, for a state with $\Bra{\vec n;\vec m}$, its T-dual includes the exchange of 
$\vec n$ and $\vec m$
as well as the extra phase as in (\ref{zeromode dual}).
}.
For the conjugate state (\ref{RRstate}), the T-dual of the direction $x^1$ and $x^2$,
which is relevant to our analysis, is given by
\be
\Bra{\CA'}=\Bra{\CA}{\cal T}_1^\dagger{\cal T}_2^\dagger
=\Bra{[C]}\sum A_{\mu_1\cdots\mu_q}\theta^{\mu_1}\cdots\theta^{\mu_q}
{\cal T}_1^\dagger{\cal T}_2^\dagger~.
\ee
It is straightforward to evaluate: 
\bea
\Bra{\CA}{\cal T}_1^\dagger{\cal T}_2^\dagger
&=&\Bra{[C]}(\CA^{(0)}+\CA^{(1)}_1\theta^1+\CA^{(1)}_2\theta^2+\CA^{(2)}_{12}\theta^1\theta^2){\cal T}_1^\dagger{\cal T}_2^\dagger\cr
&=&\Bra{[C]}(\CA^{(0)}\theta^1\theta^2-\CA^{(1)}_1\theta^2+\CA^{(1)}_2\theta^1-\CA^{(2)}_{12})~,
\label{dual RRstate}
\eea
where we have expanded the sum of RR-potentials $\CA$ in terms of $\theta^1$ and $\theta^2$.
Note that the coefficients $\CA^{(k)}$ do not contain $\theta^1$ or $\theta^2$.
From this we get the T-duality rule for the RR antisymmetric field as
\be
\CA'^{(0)}=-\CA^{(2)}_{12}~,~\CA'^{(1)}_1=\CA^{(1)}_2~,~\CA'^{(1)}_2=-\CA^{(1)}_1~,~\CA'^{(2)}_{12}=\CA^{(0)}~~.
\label{Fukuma rule}
\ee
This is essentially the Buscher rule for RR-potentials \cite{RR Buscher}
as argued in \cite{Zumino,Fukuma,Hassan}.
The overall sign is again ambiguous, depending on the convention, but the relative sign
is always present.

We will now show that
the coupling of the RR-potentials to the $D$-brane
\be
I_{CS}=\Bra{\CA}[GSO]c_0\tilde c_0\Ket{\CB}~~,\label{CSterm}
\ee
is invariant under the T-duality.
$\Ket{\CB}$ denotes the full (bosonic plus fermionic) boundary state,
including the Wilson loop factor.
By combining the T-duality rule for RR-potentials (\ref{dual RRstate}) and 
the boundary state (\ref{kD2-ND0}), it is straightforward to show this. 
Furthermore,
the invariance of the Chern-Simons coupling follows rather trivially 
from our construction.

Note first that 
the non-zero bosonic and fermionic modes do not contribute to the RR-coupling
(\ref{CSterm}).
Then, if we concentrate on the $\Bra{0;0}$ state for RR-potential, 
the T-duality transformation of the zeromode part of the corresponding amplitude 
is performed by applying the T-duality operator ${\cal T}=\Pi_\alpha{\cal T}_\alpha$, 
where
$\alpha$ are the directions of the T-duality transformation.
$I_{CS}$ is trivially rewritten by inserting ${\cal T}^\dagger{\cal T}=1$:
\be
I_{CS}=\braket{\CA|[GSO]{\cal T}^\dagger{\cal T} c_0\tilde c_0|\CB}~.\label{CSterm2}
\ee 
Note that when ${\cal T}_\alpha$ commutes with $[GSO]$, 
the parameter $p$ in the (\ref{GSOproj}) is changed to $p+1$ 
giving the corresponding GSO projection for the 
T-dual theory. 
Taking this twice into account, we can rewrite (\ref{CSterm2}) as
\be
I_{CS}=\braket{\CA|{\cal T}^\dagger[GSO] c_0\tilde c_0{\cal T}|\CB}
=\braket{\CA'|[GSO] c_0\tilde c_0|\CB'}~.
\ee
The last expression is the RR-coupling after the T-duality transformation,
written by (\ref{dual RRstate}) and (\ref{kD2-ND0}).
Therefore, we see that the Chern-Simons term is invariant under the T-duality transformation%
\footnote{ 
For the RR-state with fixed non-zero bosonic zeromodes $\Bra{\vec n;\vec m}$,
(\ref{CSterm}) is not invariant, but the T-dual gives the coupling to $\Bra{\vec m;\vec n}$.
The phase factor does not contribute to it.}.
This resolves the puzzle on the sign raised in the discussion above.
Starting from $ND2/kD0$-branes, 
both the boundary state and the RR-potential require a relative sign 
through the T-duality,
and consequently the RR-charges are kept positive to give $kD2/ND0$-branes. 
It is instructive to see this again in terms of low energy effective theory, in the 
following.

It is known that the RR-coupling (\ref{CSterm}) 
can be represented in terms of differential forms \cite{Green,Minasian,Sundell} as
\be
I_{CS}=\mu_2 \int_{M} \CA\wedge {\rm Tr}_N (e^{2\pi\alpha' F})~,
\label{true CSterm}
\ee
where $\mu_2=T_2$ is the unit of $D2$-brane charge, $M =\real \times T^2$ is the 
worldvolume, $\CA$ is a sum of 
RR-potentials (odd forms in type IIA) and $F$ is curvature $2$-form.
This is obtained
by identifying the operator $\theta^\mu$ with the differential form $dx^\mu$, 
and taking the bosonic zeromodes into account to give the worldvolume integral
 (for a derivation, see for example \cite{9707068,AST} and also Appendix.). 
In our gauge configuration (\ref{gaugeconfiguration}), this reduces to
\bea
I_{CS}&=& \mu_2 \int_{M}(\CA^{(0)}+\cdots+\CA^{(2)}_{12}dx^1\wedge dx^2)
(N+k dx^1\wedge dx^2)\cr
&=&N\mu_2 \int_M \CA^{(2)}_{12}dx^1\wedge dx^2 + k\mu_0 \int_{\real} \CA^{(0)},
\label{true CSterm2}
\eea
where $\mu_0=(2\pi \sqrt{\alpha'})^2 \mu_2$ is the unit of $D0$-brane charge, 
and we have used the same notation for the RR $q$-form field as (\ref{dual RRstate}).
Of course, only the $3$-form $A_{012}$ and the $1$-form $A_0$ survive in this expression,
and this gives the correct RR-coupling for $ND2/kD0$-branes.

The transformation rule (\ref{Fukuma rule}) can be represented 
in terms of differential forms in a compact form as
\be
\CA'=-\int_{T^2}\CA e^{dx^i\wedge dy_i}~~,
\label{Hori}
\ee
where $\CA'$ is the T-dual RR antisymmetric field with $\theta^1$, $\theta^2$ being 
replaced by $dy^1,dy^2$ and $\theta^k$ ($k\not=1,2$) replaced by $dx^k$, respectively.
Up to an overall sign, this expression of the T-duality rule for RR $q$-form field 
is known as the "Hori formula" \cite{Hori}. 

On the other hand, the term ${\rm Tr}_N(e^{2\pi\alpha' F})$ in the Chern-Simons term 
(\ref{true CSterm}) is the string theory counterpart of the Chern character ${\rm ch}(E)$. 
Since we have shown the consistency between T-duality and Nahm transformation,
its T-dual is also written similar to the family index formula (\ref{ASind}) as
\bea
{\rm Tr}_k(e^{2\pi\alpha' \tilde{F}})&=&
\frac{1}{(2\pi)^2 \alpha'}\int_{T^2} e^{dx^i\wedge dy_i}
\wedge {\rm Tr}_N(e^{2\pi\alpha' F})
\cr&=&k -N dy^1\wedge dy^2,
\label{modified ASind}
\eea
where $\tilde{F}$ is the dual curvature $2$-form corresponding 
to (\ref{dual gauge config}).
Note that the Chern character of the Poincar\'e bundle ${\rm ch}(P)$ in (\ref{ASind}) 
appears in both (\ref{Hori}) and (\ref{modified ASind}) with a slight modification.

By using these two transformations (\ref{Hori}) and (\ref{modified ASind}), we obtain
\bea
\mu_2 \int_{\tilde{M}} \CA'\wedge {\rm Tr}_k(e^{2\pi\alpha' \tilde{F}})
&=&\mu_2 \int_{\tilde{M}}(\CA'^{(0)}+\cdots +\CA'^{(2)}_{12}dy^1\wedge dy^2
)(k-N dy^1\wedge dy^2)
\cr&=&-\mu_2\int(-N\CA^{(2)}_{12} dy^1\wedge dy^2-k\CA^{(0)}dy^1\wedge dy^2)
\cr&=&N\mu_2 \int_{\tilde{M}}\CA^{(2)}_{12}dy^1 \wedge dy^2  +k\mu_0 \int_{\real} \CA^{(0)},
\eea
where $\tilde{M}=\real\times \tilde{T}^2$.
This shows the invariance of the Chern-Simons term, as required.
As we have already remarked, each transformation rule of 
RR-potentials (\ref{Hori}) or 
the gauge flux (\ref{modified ASind}) is ambiguous in its overall sign.
It is necessary in our convention that (\ref{Hori}) has a overall minus sign
to obtain a consistent and T-duality invariant Chern-Simons coupling of 
the brane and $q$-form field.

\section{Conclusion and Discussion}

We have investigated the compatibility of various T-duality rules represented 
in various forms, in the case of the $ND2/kD0$-bound state on a torus $T^2$.

First, we described in detail the two dimensional version of the Nahm transformation,
which interchanges the rank $N$ of the gauge group and the flux $k$ as 
$(N,k)~\rightarrow~(k,-N)$ together with the map $T^2$ to $\tilde T^2$,
emphasizing the relative sign.

We have then proved the equivalence of the Nahm transformation with the T-duality 
transformation in superstring theory by constructing the consistent extension of 
the boundary state description of magnetized D-branes 
on tori to the superstring.

Here, 
the T-duality transformation of the RR-zeromode sector has to be considered carefully.
We followed the construction of the boundary state given in {\cite{Billo,9707068}, 
except for introducing a new 
representation of the zeromodes.
With this representation the boundary state and the RR-states are 
treated in a seamless way,
and in particular, the T-duality 
invariance of the Chern-Simons term becomes transparent.
We also introduced the T-duality operator for the zeromode part which acts both on the boundary state and the RR $q$-form state,
that was first introduced to describe the Buscher rule of 
RR-potentials in \cite{Zumino,Fukuma,Hassan}.
This clarifies the relationship between T-duality rule at the superstring level,
and that at the low energy effective theory.
The so-called 'Hori formula' \cite{Hori} has also been derived including a consistent sign
factor.
As a result, we showed the compatibility among the T-duality, Buscher rule (Hori formula, (\ref{Hori}))
and the Nahm transform (family index formula, (\ref{modified ASind})).
\\

The overall sign appearing in these formulae indicates
a $\integer_4$-duality nature of the T-duality.
It is a well known fact that the square of the Fourier transformation, i.e., performing 
two Fourier transformations in sequence, does not give back the original function 
but the function where the variable $x$ is replaced by $-x$, the "parity transformed function".
It means $(\vec x,\vec p)\to (\vec p,-\vec x)\to (-\vec x,-\vec p)$.
To get back to the original function, one must perform the 
Fourier transform 4 times, called $\integer_4$-duality.

The same property holds for the Nahm transformation, 
which is a special case of Fourier-Mukai transformation.
Symbolically, we can write $(N,k)\to(k,-N)\to -(N,k)$.
As remarked in \S 2, the second Nahm transformation is the inverse one, 
i.e., the Dirac zero modes are left-handed spinors.
The minus sign in front of the third item denotes the overall minus sign in front of 
the family index formula. 
Similarly, RR-potentials get the overall minus sign when we transform 
them twice by the Hori formula.
These are simply the consequence of the square of the T-duality.
In fact, for ${\cal T}={\cal T}_2{\cal T}_2$ we have
${\cal T}^2=-{\cal T}_2^2{\cal T}_1^2=-1$.

On the other hand, T-duality is usually designed to be a $\integer_2$-duality \cite{Hassan}.
This is possible because we can redefine both RR-potentials and the boundary state
by a minus sign using the sign ambiguity.  
But we emphasize that they have to be simultaneously redefined 
so that their overlap is unchanged.
 \\

It is interesting to extend our analysis to a higher even-dimensional torus $T^{d}$ ($d$: even),
where the Nahm transformation interchanges the rank and higher Chern numbers.
For example, there are so-called toron solutions on $T^4$ corresponding to 
$D4/D2/D0$-bound states \cite{Rangoolam}.

The discrete T-duality considered in this paper is a part of the full T-duality 
group $O(2,2;\integer)$ for $T^2$ and  $O(d,d;\integer)$ for $T^d$.
Corresponding T-duality rules for bosonic boundary states \cite{Duo,DiVecc}, 
and for RR-potentials \cite{Zumino,Fukuma,Hassan} are known.
This suggest that there is an $O(d,d;\integer)$-family of Nahm transformations.
However, for $d=2$, 
this family includes not only $D2/D0$-bound states but also tilted $D$-strings 
and a state of $D0$-branes only.

Even if restricting to type IIA theory and the subgroup $SO(d,d;\integer)$,
the Nahm transformation should relate gauge theories of various (even) dimensions.
This in one way leads to the derived category viewpoint on $D$-brane bound states, 
and the corresponding Nahm transformation would be the Fourier-Mukai transformation.
Another possibility is to use scalar fields on lower dimensional $D$-branes.
For example, 
$D2/D0$-bound states described either by $D2$-branes with non-trivial gauge bundles on $T^2$,
or by $D0$-branes with non-trivial scalar configurations. 
The latter picture is widely used in constructing gauge theories on the dual torus 
in terms of matrix models of $D0$-branes, but the complete description in terms of 
boundary states is not clear, as opposed to the $\real^{2}$ case \cite{Ishibashi}.
There are some previous attempts \cite{Kato,Youm} on this subject for a torus.
Once it is known, the connection of the T-duality and the Morita equivalence 
becomes more transparent.

An odd sequence of discrete T-dualities is also interesting, because 
it relates type IIA and type IIB theory, as mirror symmetry. 
In order treat all RR-states as a single $O(d,d;\integer)$ multiplet, 
it is important to perform the transformation before GSO projection.
In this sense, we believe that our new representation of RR-zeromodes 
serves as suitable basis for further study on such cases.\\

\quad\\
{\large\bf Acknowledgment}\\
The authors would like to thank M. Hamanaka, H. Kajiura, M. Fukuma, M. Kato,
N. Yokoi, S. Sasa, and S. Ramgoolam for discussions and valuable comments.
T. A. is supported by the GCOE program, ``Weaving Science Web beyond Particle-Matter Hierarchy."

\appendix
\section{Appendix}
\subsection{Fermionic part of the boundary state}

The fermionic part of the boundary state has been formulated in 
\cite{Callan,PolCai88,Billo,9707068}.
Since we introduced a new representation for the zeromode part, we summarize the construction of the relevant parts of the boundary state here. 

The basic OPE of the fermionic field on the world sheet and spin operators are 
specified by the 10 dimensional gamma matrices $\Gamma^\mu$. 
We take the following choice
\be
\psi^\mu S_{-\half A}\sim{1\over\sqrt{2}}\Gamma^\mu_{AB}S_{-\half}^B
~,~~\tilde \psi^\mu \tilde S_{-{3\over2}A}
\sim{1\over\sqrt{2}}(i\Gamma_{11}\Gamma^\mu)_{AB}\tilde S_{-{3\over2}}^B~,
\ee
where rightmovers are distinguished from the leftmovers by a tilde. 
Here, we introduced the factor $i\Gamma_{11}$ for the rightmovers. This convention is 
different from the representation given in \cite{Billo}, but it has the advantage that 
it simplifies the definition of the boundary states and it provides a seamless 
treatment of the RR-antisymmetric tensor field state and the boundary state. 

Corresponding to the spin operators, we define the states $\Ket{A,-1/2}$ and $\Ket{\tilde B,-3/2}$  with spinor index $A, B$ and pictures $-\half, -{3\over2}$, respectively\footnote{We distinguish the rightmover by the tilde on the index.}. 
The RR-states of the closed string, on which we construct the boundary state, is asymmetric in the picture and given by the 
tensor product 
\be
\Ket{A,-1/2}\otimes\Ket{\tilde B,-3/2}\ .
\ee

The zero modes of the fermion fields in the RR sector 
are represented by the 
Dirac matrices on these states
\be
\psi_0^\mu={1\over\sqrt{2}}(\Gamma^\mu\otimes\id)~,~~
\tilde \psi_0^\mu={1\over\sqrt{2}}(\Gamma_{11}\otimes i\Gamma_{11}\Gamma^\mu)~.
\label{Srepresentation}
\ee
The vacuum of the RR-sector which is an element of BRST infinite cohomology \cite{Billo,Berkovits} with an asymmetric background charge is then
\be
\Ket{[C]}=e^{i\gamma_0\tilde\beta_0}
\Ket{A,-1/2}C_{AB}\Ket{\tilde B,-3/2}\label{RRvacuum}~,
\ee
where $C_{AB}$ is the charge conjugation matrix and $\gamma_0$ and $\beta_0$ are the zeromode of the $\beta\gamma$ ghost in superstring. (We will suppress the tensor product in the following.)

The zeromode part of the boundary state can be also
constructed by the linear combination of the above product states
 represented by using a matrix $M_\eta'^A{}_B$ as
\be
\Ket{[CM'_\eta],\eta}=e^{i\eta\gamma_0\tilde\beta_0}
\Ket{A,-1/2}[CM'_\eta]_{AB}\Ket{\tilde B,-3/2}~.
\ee
Here we introduced a parameter $\eta=\pm$ to distinguish the boundary conditions.
The Dirichlet condition along $\mu$-direction means to impose
(Dirichlet condition for $\eta=1$ is the same as the Neumann condition 
for $\eta=-1$ and vice versa, i.e., $\Ket{D,\eta}=\Ket{N,-\eta}$.) 
\be
(\psi_0^\mu+i\eta\tilde \psi_0^\mu) \Ket{[CM'_\eta],\eta}=0~,
\ee
and the corresponding condition for the ghosts.
Using (\ref{Srepresentation}), we get the condition 
for the matrix $M'_\eta$ as
\be
\Ket{[C(\Gamma^\mu M'_\eta-\eta \Gamma_{11}M'_\eta\Gamma_{11}\Gamma^\mu)],\eta}=0~.\label{Dcondition}
\ee
For the D-instanton, the condition (\ref{Dcondition}) must be satisfied for all $\mu$,
and we obtain 
\be
M'_\eta=\bigg\{\matrix{ccl}{1 &\mbox{\rm for}& \eta=1\cr \Gamma_{11}&\mbox{\rm for}&\eta=-1
}.\ee 
In the same way we can obtain the matrix $M'_\eta$ for the boundary state of a $Dp$-brane. However, it is easier to use the operators $\theta^\mu$ introduced in (\ref{thetaoperators}).
For a D$p$-brane the boundary state 
of the RR sector is proportional to \cite{9707068}
\be
\Ket{Dp,\eta}=e^{i\eta\gamma_0\tilde\beta_0}\Ket{A,-1/2}[CM_\eta'\Gamma^{01\cdots p}]_{AB}\Ket{\tilde B,-3/2}\ . \label{Dpbrane}
\ee
It can be constructed by introducing the following operator:
\be
\theta_\eta^\mu={1\over\sqrt{2}}(\psi_0^\mu+i\eta \tilde \psi_0^\mu)~.
\ee
Note that
\be
\{\theta_\eta^\mu,\theta_\eta^{\nu\dagger}\}=\half(g^{\mu\nu}+\eta^2 g^{\mu\nu})=g^{\mu\nu}~.
\ee
The zeromode part of the boundary state of the D$p$-brane is characterized by
\bea
\theta^{\mu\dagger}_\eta\Ket{Dp,\eta}=0~~&\mbox{for}&~~\mu=0,\cdots,p~(\mbox{Neumann}),\cr
\theta^\mu_\eta\Ket{Dp,\eta}=0~~&\mbox{for}&~~\mu=p+1,\cdots,9~(\mbox{Dirichlet}),
\eea
and obtained systematically by applying $\theta_\eta^\dagger$ on the $D$ instanton state:
\be
\theta_\eta^{0\dagger}\cdots\theta_\eta^{p\dagger}\Ket{D(-1),\eta}
=(-\eta)^{p+1}\Ket{Dp,\eta}~.
\ee
Note that the RR vacuum defined in (\ref{RRvacuum}) using the representation (\ref{Srepresentation}) is equivalent to the zeromode part of the
 boundary state of the D-instanton $\Ket{D(-1),+}$.

The coupling of the brane to the RR-antisymmetric tensor fields 
can be obtained by taking the pairing of the state of the RR-antisymmetric tensor field with the boundary state. Since the picture of the boundary state is taken to be asymmetric, the RR state should be also prepared in the asymmetric picture \cite{Billo,Yost}.

\subsection{Pairing of RR-state and Boundary state}

The overlap given in (\ref{CSterm}) is the pairing of the RR $p$-form state $\Bra{\CA}$ and the boundary state $\Ket{\CB}$. However, this expression has divergences. Following
Billo et al.\cite{Billo}, we regularize it by the regulator $x^{2(F_0+G_0)}$ and take the limit $x\rightarrow 1$ in the end. Thus we need to evaluate 
\be
I_{CS,x}=\braket{[CN]|x^{2(F_0+G_0)}[GSO]c_0\tilde c_0|[CM]}~,
\ee
where $M$ and $N$ are products of $\Gamma$ matrices.
Considering the ordering of the conjugate state in our convention, we define the 
inner product for the spinor states as 
\be
(\Bra{\tilde A}\Bra{B})(\Ket{C}\Ket{\tilde D})=(C^{-1})^{BC}(C^{-1})^{AD}.
\ee
Then one can derive 
\be
(\Bra{\tilde A}[CN]_{AB}\Bra{B})(\Ket{C}[CM]_{CD}\Ket{\tilde D})
=\Tr{CNC^{-1}CMC^{t-1}}=-\Tr{NM}.
\ee
From the $\beta\gamma$-ghosts we get \cite{Billo}
\be
\Bra{-\tilde\half,-{3\over2}}e^{i\eta_1\beta_0\tilde\gamma_0}x^{-2G_0}e^{i\eta_2\gamma_0\tilde \beta_0}\Ket{-\half,-\tilde{3\over2}}={x\over 1-\eta_1\eta_2x^2}.
\ee
The GSO projection operator is given in (\ref{GSOproj})
\be
[GSO]\Ket{[CM],+}
={1\over4}\sum_{\eta=\pm}\Ket{[CM'_\eta(M-(-1)^p\Gamma_{11}M\Gamma_{11})],\eta}~.
\ee
The pairing of the boundary state with the states of the RR-potential (\ref{RRstate}) gives terms like
\be
\Bra{[CN]}x^{2F_0+2G_0}\Ket{[CM'_\eta K],\eta}=-{x\over 1+\eta x^2}
\Tr{N(x^{-2|A|}M'_\eta) K},
\ee
where $K=M-(-1)^p\Gamma_{11}M\Gamma_{11}$. 
The matrix $[x^{-2|A|}M'_\eta]^A{}_{B}$ is a diagonal matrix where $|A|$ is 
the sum of the weight when the spinor state $\Ket{A}$ is labeled by the
 weight of the spinor representation. 
For the $D2D0$ state, we have $K=2(N\Gamma_{210}+k\Gamma_0)$.

\end{document}